\documentclass{article}
\usepackage{spconf,amsmath,graphicx,hyperref}
\usepackage{tabularx}
\usepackage{booktabs}
\usepackage{adjustbox}
\usepackage{amsfonts}
\usepackage{afterpage}
\usepackage{tablefootnote}
\usepackage{hyperref}

\newcommand{\std}[1]{\textsuperscript{\scriptsize\,±#1}}

\title{EdgeSpot: Efficient and High-Performance Few-Shot Model for Keyword Spotting}
\name{Oguzhan Buyuksolak, Alican Gok, Osman Erman Okman}
\address{Analog Devices, Istanbul, Turkey}

\begin{document}
%\ninept
%
\maketitle

\begingroup
\renewcommand\thefootnote{}
\footnotemark\footnotetext{
\textcopyright~ 2026 IEEE. Personal use of this material is permitted. Permission from IEEE must be obtained for all other uses, in any current or future media, including reprinting/republishing this material for advertising or promotional purposes, creating new collective works, for resale or redistribution to servers or lists, or reuse of any copyrighted component of this work in other works.

To appear in Proceedings of the IEEE International Conference on Acoustics, Speech, and Signal Processing (ICASSP), 2026.}
\endgroup

\thispagestyle{empty}
\pagestyle{empty}

\begin{abstract}
We introduce an efficient few-shot keyword spotting model for edge devices, EdgeSpot, that pairs an optimized version of a BC-ResNet-based acoustic backbone with a trainable Per-Channel Energy Normalization frontend and lightweight temporal self-attention. Knowledge distillation is utilized during training by employing a self-supervised teacher model, optimized with Sub-center ArcFace loss. This study demonstrates that the EdgeSpot model consistently provides better accuracy at a fixed false-alarm rate (FAR) than strong BC-ResNet baselines. The largest variant, EdgeSpot-4, improves the 10-shot accuracy at 1\% FAR from 73.7\% to 82.0\%, which requires only 29.4M MACs with 128k parameters.

\end{abstract}

\section{Introduction}
\label{sec:intro}
Keyword Spotting (KWS) enables hands-free interaction by detecting specific spoken commands or wake-words in audio, but conventional neural network (NN) based systems require extensive data and significant computational resources during training. Adapting the model for different sets of keywords requires repeating the same training procedure; therefore, it is not feasible to adapt the models on edge hardware. Few-shot keyword spotting (FS-KWS) overcomes these limitations by using deep neural networks with compact architectures and metric learning techniques \cite{FewShotMazumderIntro}, such as prototypical networks, to recognize new keywords from just a few examples. These systems learn an embedding space where similar keywords cluster together, allowing users to enroll custom keywords with minimal data.
In open-set scenarios, users enroll new keywords by providing a small number of spoken samples, which are generally averaged in the embedding space to form class prototypes \cite{UnsupervisedFewShotKWS}. During inference, the system classifies test samples by measuring their proximity to these prototypes. 

In \cite{RusciMicro, RusciInterspeech}, a ResNet~\cite{Resnet} based model is trained with triplet loss to achieve good FS-KWS application performance. Recently, another training scheme was suggested that leverages Self-supervised learning (SSL) models for robust feature extraction \cite{Our_FSKWS}. A teacher–student framework for knowledge distillation is proposed to train the same ResNet architecture from the reduced dimensionality embeddings of the pretrained SSL model. Still, this model architecture is not optimized enough for edge deployment.

However, in the fixed vocabulary KWS literature, there are plenty of options for efficient model architectures~\cite{MatchboxNet, DSResNet, TCResNet}. Among them, BC-ResNet ~\cite{BCResnet} has emerged as a state-of-the-art model on the Google Speech Commands (GSC)~\cite{GSC} benchmark, while also offering computational efficiency and scalability. Motivated by the strong efficiency–accuracy trade-offs of this model architecture, we introduce EdgeSpot for the few-shot setting: a compact, edge-friendly model. EdgeSpot augments the BC-ResNet backbone with a trainable Per-Channel Energy Normalization (PCEN)~\cite{PCEN} frontend and lightweight temporal self-attention to yield more discriminative embeddings at low false-alarm rates (FAR).

In this work, we provide a comprehensive overview of the proposed architecture, detailing its structure as well as its performance in terms of accuracy and computational efficiency. To ensure a fair comparison, we adapt BC-ResNet to the same FS-KWS training method and prototype-based inference protocol. Furthermore, we analyze the enhancements that EdgeSpot introduces over the baseline ResNet architecture described in \cite{Our_FSKWS}. Through side-by-side evaluations, we demonstrate that EdgeSpot consistently outperforms BC-ResNet and the baseline on both the Multilingual Spoken Words Corpus (MSWC)~\cite{MSWCdataset} and cross-domain GSC datasets, all while maintaining the low computational and parameter requirements essential for on-device deployment.

\section{PROPOSED METHOD}
\label{sec:method}

\subsection{Model Architecture}
\begin{figure*}
\centerline{\includegraphics[width=\textwidth]{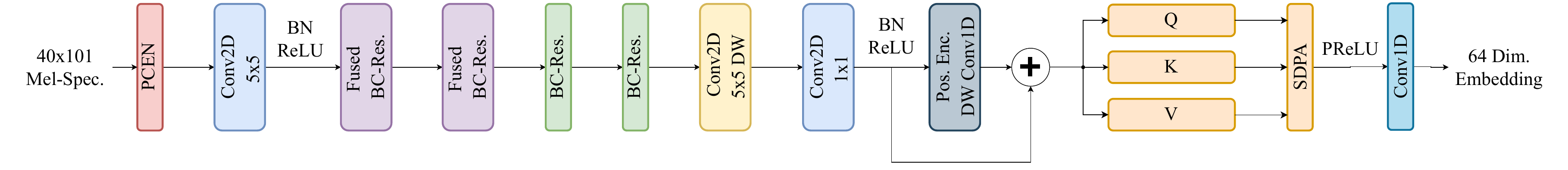}}
\caption{Architecture of the Proposed Model. The model processes a 40×101 Mel-Spectrogram input and applies Per-Channel Energy Normalization (PCEN) for preprocessing. It then passes through a 5×5 Conv2D layer followed by depthwise 2D convolutions (DW Conv2D) and residual blocks (BC-Res. and Fused BC-Res.) to extract hierarchical features. Positional encoding is incorporated using a depthwise Conv1D layer to enhance temporal feature representation. A self-attention mechanism (SDPA) focuses on capturing dependencies in the time dimension. Finally, a Conv1D layer reduces dimensionality, producing a 64-dimensional embedding as the output.}
\label{fig:arch}
\end{figure*}
EdgeSpot adopts BC-ResNet as its acoustic backbone and introduces three targeted, edge-friendly improvements that raise accuracy while preserving efficiency: (a) an additional trainable PCEN frontend before the acoustic stack, (b) an early-block temporal-path fusion that replaces the depthwise separable convolution with a single temporal convolution, and (c) insertion of a lightweight temporal self-attention head. Fig.~\ref{fig:arch} and Table~\ref{tab:table1} present the details of the EdgeSpot architecture. Details are explained in the following sections.

\begin{table}[ht]
\small % or \footnotesize
\centering
\caption{Details of the model architecture for $\tau = 1$, where $\tau$ is a multiplier that determines the output channel width. The final \texttt{conv1d} layer always has an output channel of 1, regardless of the value of $\tau$.}
\label{tab:table1}
\begin{tabularx}{\columnwidth}{@{}lXcccc@{}}
\toprule
\textbf{Input} & \textbf{Operator} & \textbf{n} & \textbf{c} & \textbf{s} & \textbf{d} \\
\midrule
1$\times$40$\times$101  & PCEN                         & -- & -- & --     & --     \\
1$\times$40$\times$101  & conv2d(5$\times$5)-BN-ReLU   & -- & 16 & (2,1) & 1      \\
16$\times$20$\times$101 & Fused BC-ResBlock            & 2  & 8  & 1     & 1      \\
8$\times$20$\times$101  & Fused BC-ResBlock            & 2  & 12 & (2,1) & (1,2)  \\
12$\times$10$\times$101 & BC-ResBlock                  & 4  & 16 & (2,1) & (1,4)  \\
16$\times$5$\times$101  & BC-ResBlock                  & 4  & 20 & 1     & (1,8)  \\
20$\times$5$\times$101  & DW conv2d(5$\times$5)        & -- & 20 & 1     & 1      \\
20$\times$1$\times$101  & conv2d(1$\times$1)-BN-ReLU   & -- & 32 & 1     & 1      \\
32$\times$101  & DW conv1d(16) Pos. E. & -- & 32 & 1     & 1      \\
101$\times$32  & SDPA-PReLU                   & -- & -- & --    & --     \\
101$\times$64           & conv1d(1)                    & -- & 1  & --    & --     \\
\bottomrule
\end{tabularx}
\end{table}
\subsubsection{Per-Channel Energy Normalization (PCEN)}
\label{sec:PCEN}
PCEN replaces static log compression with per-channel automatic gain control and stabilized root compression. In practice, it applies a causal IIR smoother to the mel energies and then normalizes and compresses them, as specified in Eqs.~(1)–(2) below.
\begin{equation}
\mathrm{PCEN}(t,f)=\left(\frac{E(t,f)}{(\epsilon+M(t,f))^{\alpha}}+\delta\right)^{r}-\delta^{r},
\end{equation}
\begin{equation}
M(t,f)=(1-s)\,M(t-1,f)+s\,E(t,f).
\end{equation}

Here, $E(t,f)$ is the mel-filterbank energy; $M(t,f)$ is a causal per-channel IIR smoother with $s\in(0,1)$; $\epsilon>0$ is a small constant; $\alpha\in[0,1]$ controls AGC strength; and $\delta>0$, $r\in(0,1]$ define the stabilized root compression \cite{PCEN,PCEN_Why}. All operations are differentiable, enabling end-to-end learning \cite{PCEN}.

PCEN reduces loudness dependence, suppresses stationary backgrounds, and enhances onsets; it is widely reported to improve robustness in noisy and far-field speech. Moreover, it tends to Gaussianize magnitude distributions and decorrelate frequency bands, which benefits downstream discriminative models and cross-domain transfer \cite{PCEN,PCEN_Why}. We include PCEN in EdgeSpot primarily to strengthen cross-domain generalization. In initial ablations, adding only PCEN with no other architectural change improved cross-domain test set performance (GSC, see section~\ref{training meth.}), which we attribute to PCEN’s ability to Gaussianize and whiten spectrogram magnitudes~\cite{PCEN_Why}.

In EdgeSpot, we place a PCEN layer before the acoustic backbone and train it jointly with the model, learning channel-shared scalars $(\alpha, r, \delta)$ and a channel-shared smoothing coefficient $s$, while fixing $\epsilon=10^{-6}$.
\subsubsection{Fusion of Early Blocks}
\label{sec:fusion}
\begin{figure}[htb]
  \centering
  \centerline{\includegraphics[width=\columnwidth]{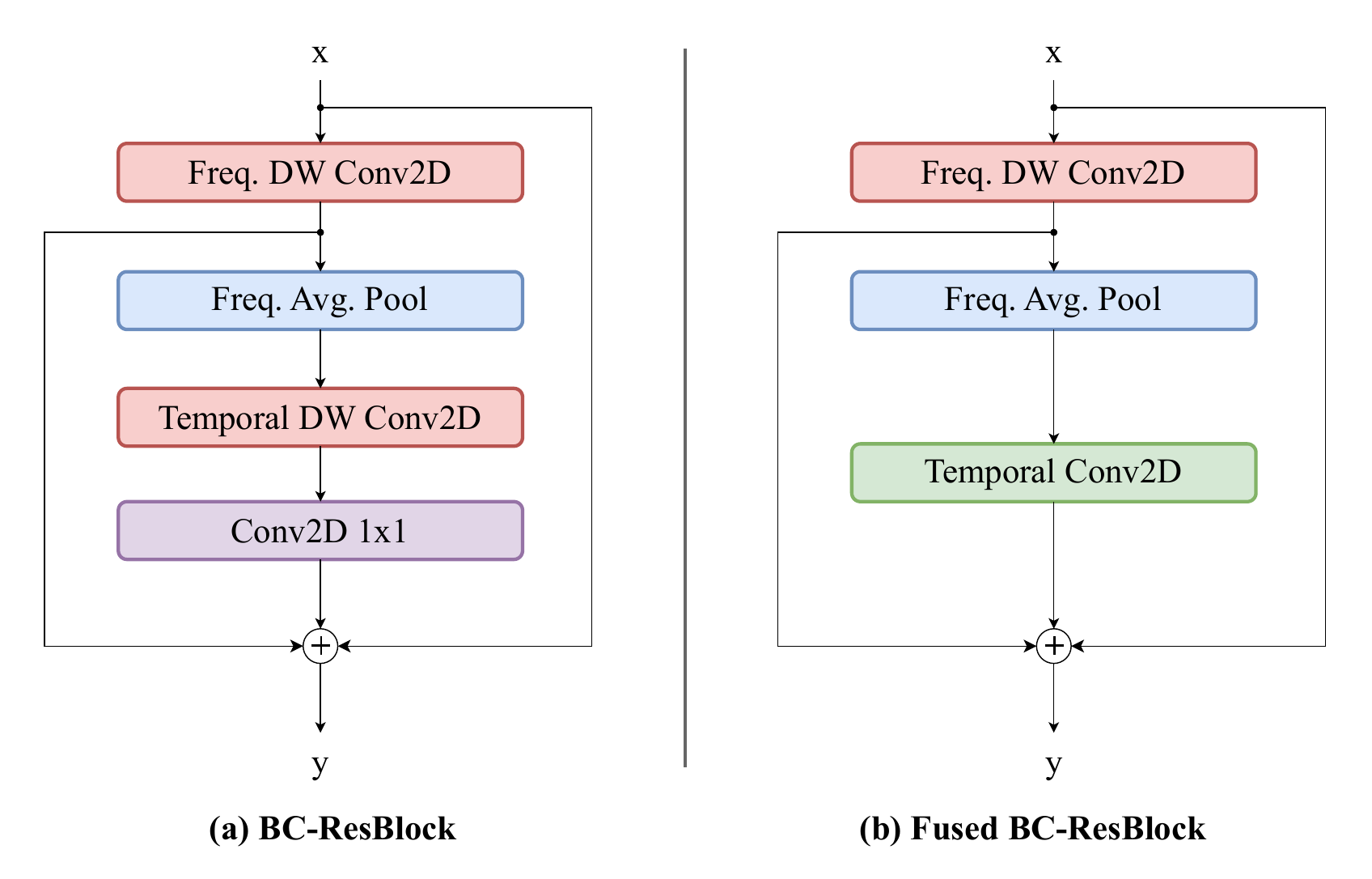}}
%  \vspace{2.0cm}
\caption{Comparison of (a) BC-ResBlock and (b) Fused BC-ResBlock, highlighting the replacement of 'Temporal DW Conv. + 1x1 Conv.' with a single 'Temporal Conv.'.}
\label{fig:bcres}
\end{figure}

 Fig.~\ref{fig:bcres} (a) presents the BC-ResBlock as given in \cite{BCResnet}. The first layer is a 3$\times$1 frequency-depthwise convolution with SubSpectral Normalization, which averages the activation along the frequency dimension. It is followed by a temporal depthwise convolution and then by a 1$\times$1 pointwise convolution to broadcast the temporal feature back to 2D. Lastly, an auxiliary 2D residual connection is added to form the output of the block.

In our architecture, for the audio backbone, we adopt the same early-fusion principle with EfficientNetV2 \cite{efficientnetv2}, which shows replacing depthwise-separable stacks with a single regular convolution is beneficial in the earliest stages. Therefore, we first define a Fused BC-ResBlock, which replaces the last two layers of BC-ResBlock with a single regular temporal convolution as seen in Fig.~\ref{fig:bcres}(b). Note that the new convolution operation is configured with the same kernel, stride, and dilation and with output channels matched to the original 1$\times$1 projection, while leaving the frequency branch, frequency averaging and broadcasting, the auxiliary 2D residual, and the transition-block behavior unchanged.

We apply this fused temporal path only in the earliest stages (see Fig.~\ref{fig:arch}) to capture the accuracy benefits observed for early-stage fusion in EfficientNetV2 while avoiding the regressions seen when fully fusing all stages~\cite{efficientnetv2}. In our setting, the incremental parameters and operations are minimal, since early stages have small channel counts and the fused operation is 1D on the temporal branch, yet the simplification of the separable stack improves optimization in practice.

\subsubsection{Relative Positional Encoding}
\label{sec:rpe}
\afterpage{%
\begin{table*}[ht!]
\centering
\caption{Model performance on MSWC (1-/10- shot) and GSC (1-/10- shot). Each cell reports mean$_{1\text{-shot}}$/mean$_{10\text{-shot}}$ with std as a tiny superscript.}
\label{tab:edge_mswc_gsc_shots}

% compact spacing
\setlength{\tabcolsep}{6pt}
\renewcommand{\arraystretch}{1.1}

\small
\begin{adjustbox}{max width=\textwidth}
\begin{tabular}{l|ccc|ccc|c|c}
\toprule
\textbf{Model} &
\multicolumn{3}{c|}{\textbf{MSWC}} &
\multicolumn{3}{c|}{\textbf{GSC}} &
\textbf{\#MACs} & \textbf{\#Params} \\
& $DET_{1\%}$ & $DET_{5\%}$ & AUROC & $ACC_{1\%}$ & $ACC_{5\%}$ & AUC & & \\
\cmidrule(lr){2-4}\cmidrule(lr){5-7}
\textbf{Shots} & \multicolumn{3}{c|}{\scriptsize 1-/10- shot} & \multicolumn{3}{c|}{\scriptsize 1-/10- shot} & & \\
\midrule
Teacher~\cite{Our_FSKWS} &
91.4\std{0.03}/97.0\std{0.09} &
97.4\std{0.02}/99.3\std{0.05} &
99.3\std{0.01}/99.8\std{0.01} &
65.1\std{7.6}/83.0\std{0.5} &
75.1\std{6.9}/85.2\std{0.5} &
82.7\std{5.7}/90.9\std{0.7} &
63.3\text{*} G & 217.8\text{*} M \\
\midrule
ResNet15~\cite{Our_FSKWS} &
86.3\std{0.13}/96.1\std{0.05} &
93.9\std{0.03}/98.4\std{0.02} &
98.3\std{0.01}/99.5\std{0.01} &
44.3\std{7.4}/75.4\std{1.4} &
60.6\std{8.0}/85.2\std{0.8} &
73.7\std{6.9}/91.0\std{0.5} &
235 M & 480 k \\
\midrule
BC-ResNet-1 &
61.2\std{0.08}/81.2\std{0.05} &
85.3\std{0.08}/95.5\std{0.13} &
97.0\std{0.01}/98.9\std{0.01} &
14.1\std{3.9}/35.6\std{2.7} &
27.3\std{5.6}/52.9\std{2.5} &
48.5\std{6.6}/71.7\std{1.7} &
2.5 M & 10.9 k \\
Edgespot-1  &
\textbf{66.8}\std{0.13}/\textbf{85.7}\std{0.10} &
\textbf{88.9}\std{0.17}/\textbf{96.7}\std{0.01} &
\textbf{97.7}\std{0.01}/\textbf{99.2}\std{0.01} &
\textbf{17.2}\std{4.6}/\textbf{40.1}\std{2.5} &
\textbf{28.0}\std{6.0}/\textbf{54.5}\std{2.2} &
\textbf{50.8}\std{7.0}/\textbf{74.6}\std{1.4} &
4.5 M & 16.6 k \\
\midrule
BC-ResNet-2 &
77.3\std{0.17}/91.1\std{0.09} &
93.3\std{0.14}/98.2\std{0.03} &
98.5\std{0.01}/99.5\std{0.01} &
25.8\std{5.4}/53.4\std{2.5} &
42.0\std{6.8}/69.0\std{1.6} &
62.9\std{6.7}/83.1\std{0.9} &
7.3 M & 30.6 k \\
Edgespot-2  &
\textbf{79.3}\std{0.22}/\textbf{92.0}\std{0.12} &
\textbf{94.0}\std{0.08}/\textbf{98.4}\std{0.02} &
\textbf{98.6}\std{0.01}/\textbf{99.5}\std{0.01} &
\textbf{29.2}\std{5.6}/\textbf{57.1}\std{2.2} &
\textbf{48.2}\std{6.6}/\textbf{74.0}\std{1.2} &
\textbf{66.8}\std{6.2}/\textbf{85.7}\std{1.0} &
10.3 M & 43.3 k \\
\midrule
BC-ResNet-3 &
83.5\std{0.03}/94.0\std{0.18} &
95.5\std{0.05}/98.8\std{0.04} &
98.9\std{0.01}/99.6\std{0.01} &
36.8\std{7.0}/68.4\std{1.4} &
56.6\std{8.0}/79.7\std{0.8} &
\textbf{73.7}\std{7.1}/89.2\std{0.5} &
14.5 M & 59.2 k \\
Edgespot-3  &
\textbf{85.2}\std{0.19}/\textbf{94.8}\std{0.22} &
\textbf{95.9}\std{0.22}/\textbf{98.9}\std{0.04} &
\textbf{99.0}\std{0.01}/\textbf{99.7}\std{0.01} &
\textbf{41.1}\std{8.0}/\textbf{70.2}\std{1.6} &
\textbf{57.2}\std{8.3}/\textbf{80.8}\std{0.9} &
73.2\std{6.3}/\textbf{89.4}\std{0.4} &
18.6 M & 80.6 k \\
\midrule
BC-ResNet-4 &
87.1\std{0.26}/95.5\std{0.08} &
96.4\std{0.05}/\textbf{99.1}\std{0.03} &
99.1\std{0.01}/99.7\std{0.01} &
44.5\std{7.6}/73.7\std{1.2} &
60.5\std{7.8}/82.1\std{0.6} &
74.3\std{7.1}/90.6\std{0.4} &
24.1 M & 96.6 k \\
Edgespot-4  &
\textbf{87.8}\std{0.12}/\textbf{95.7}\std{0.06} &
\textbf{96.7}\std{0.04}/\textbf{99.1}\std{0.01} &
\textbf{99.2}\std{0.01}/\textbf{99.7}\std{0.01} &
\textbf{51.8}\std{8.1}/\textbf{82.0}\std{0.9} &
\textbf{67.0}\std{7.8}/\textbf{87.4}\std{0.6} &
\textbf{79.0}\std{6.6}/\textbf{91.9}\std{0.4} &
29.4 M & 128.3 k \\
\bottomrule
\end{tabular}
\end{adjustbox}
\normalsize
\begin{minipage}{\linewidth}
\footnotesize
\textsuperscript{*} The teacher model consists of a Wav2vec2 encoder up to the 16th transformer layer and the dimensionality reduction module.
\end{minipage}
\end{table*}
}
To encode temporal order without absolute indices, we use a lightweight, convolutional relative positional encoding along time. Given the channels-first sequence $X \in \mathbb{R}^{C_\tau \times T}$, we add a per-channel depthwise temporal convolution $\varphi(X)$ residually:
\begin{equation}
    \tilde{X} = X + \varphi(X), \qquad \tilde{X} \in \mathbb{R}^{C_\tau \times T}.
    \label{eq:rpe_residual}
\end{equation}
This maintains shift equivariance while allowing each channel to learn compact, local timing cues (e.g., onsets/offsets), gently conditioning features for the downstream attention without explicit $T \times T$ positional biases.

The operator $\varphi(\cdot)$ is a depthwise 1D convolution with kernel size $\kappa$ and “same” padding that preserves the output length:
\begin{equation}
    [\varphi(X)]_{c,t}
    =
    \sum_{\Delta = -\lfloor \kappa/2 \rfloor}^{\lfloor \kappa/2 \rfloor}
    K_{c,\Delta}\; X_{c,\,t+\Delta},
    \qquad c=1,\dots,C_\tau,
    \label{eq:rpe_conv}
\end{equation}
where $K \in \mathbb{R}^{C_\tau \times \kappa}$ are learnable per-channel filters and out-of-range indices are zero padded. In EdgeSpot, we set $\kappa$ = 16 and symmetric padding p = 8. The positional-enhanced sequence $\tilde{X}$ is then fed to the attention block (Sec.~\ref{SDPA}).

\subsubsection{Scaled Dot Product Attention (SDPA)}
\label{SDPA}
In EdgeSpot, we apply single-head scaled dot-product self-attention along the time axis to model long-range temporal dependencies at low cost for the short utterances considered in KWS. Let the output of the convolutional front-end be a 1D temporal sequence with $T$ frames and $C_\tau$ channels, where $C_\tau$ depends on the width multiplier $\tau$ of the network (see Fig.~\ref{fig:arch} and Table~\ref{tab:table1}). We reshape this tensor to
$
X \in \mathbb{R}^{T \times C_\tau},
$
i.e., one $C_\tau$-dimensional feature vector per time step.

We first obtain queries, keys, and values via learnable linear projections to a fixed $d=64$-dimensional space:
\begin{equation}
Q = X W_Q + \mathbf{1} b_Q^\top,\quad
K = X W_K + \mathbf{1} b_K^\top,\quad
V = X W_V + \mathbf{1} b_V^\top,
\label{eq:qkv}
\end{equation}
where $W_Q, W_K, W_V \in \mathbb{R}^{C_\tau \times d}$, $b_Q, b_K, b_V \in \mathbb{R}^{d}$, and $\mathbf{1} \in \mathbb{R}^{T}$ is an all-ones column vector that broadcasts the biases across time. In EdgeSpot, these projections map the backbone activations to $d=64$ by design, independently of $\tau$.

Scaled dot-product attention over the temporal dimension is then computed as
\begin{equation}
A = \operatorname{Softmax}\!\left(\frac{Q K^\top}{\sqrt{d}}\right) \in \mathbb{R}^{T \times T},\qquad
Z = A V \in \mathbb{R}^{T \times d},
\label{eq:sdpa}
\end{equation}
where the $\sqrt{d}$ scaling stabilizes gradients for large $d$ and the softmax is applied row-wise so that each row of $A$ forms a probability distribution over time steps. In our implementation, the attention strictly operates along the temporal axis (no spatial/frequency mixing), consistent with the 1D temporal representation before attention.

We then apply a pointwise nonlinearity to the contextualized sequence,
\begin{equation}
\tilde{Z} = \operatorname{PReLU}(Z),
\label{eq:prelu}
\end{equation}
and produce an utterance-level 64-D embedding by aggregating $\tilde{Z}$ across time with a lightweight 1D convolutional head (see Fig.~\ref{fig:arch}):
\begin{equation}
\mathbf{e} = g_\theta(\tilde{Z}) \in \mathbb{R}^{64},
\label{eq:agg}
\end{equation}
where $g_\theta$ denotes the learned temporal aggregation module with a shallow Conv1D stack in EdgeSpot. This head preserves the 64-dimensional embedding size used by our training pipeline and downstream prototype-based inference.

\subsection{Training Methodology}
\label{training meth.}
We follow the FS-KWS teacher–student pipeline from our prior work \cite{Our_FSKWS}, in which a pretrained Wav2Vec2.0 encoder~\cite{neurips2020_wav2vec_2} provides frame-level features and an attention-based dimensionality reduction head maps them to 64-D embeddings trained with Sub-center ArcFace (SCAF)~\cite{SubcenterArcFace}, forming the teacher. The EdgeSpot student is trained end-to-end by distillation from teacher embeddings with a composite objective as:
\begin{equation}
\mathcal{L} \;=\; \mathcal{L}_{\mathrm{KD}} \;+\; \lambda\,\mathcal{L}_{\mathrm{SCAF}},
\end{equation}
where $\mathcal{L}_{\mathrm{KD}}$ is the MSE between teacher and student embeddings, with $\lambda=5\times10^{-5}$. To train the networks, we use the train split of the English portion of MSWC, which contains 5.5 million 1-second samples of 39,000 unique words. We test our approach using the MSWC dataset's test split, comprised of 680k samples from 38,090 unique words, and the cross-domain GSC dataset, a collection of 100k 1-second audio clips belonging to 35 short commands. For student models, each 1-second waveform (16 kHz) is converted to a 40-band mel spectrogram using a centered STFT, yielding a $40\times101$ frequency-time map that serves as input to the acoustic front end.

A keyword prototype is formed by averaging embeddings from $K$ examples; and the decision is made by comparing the inference results against the keyword prototypes. A threshold is used to control the FAR, following the same protocol as given in \cite{Our_FSKWS}. Models are trained for 40 epochs, using the Adam optimizer with weight decay $4\times 10^{-5}$ and a cosine learning-rate schedule. The first 5 epochs linearly warm up to a peak learning rate of $1\times 10^{-3}$, after which cosine decay is applied at every optimization step. We apply SpecAugment~\cite{SpecAug} during training with settings tied to model width. For the smallest model ($\tau = 1$), no SpecAugment is used, including time-stretch. For $\tau \in \{2, 3, 4\}$, we enable SpecAugment with time-stretch factors sampled uniformly from $[0.9, 1.1]$, a frequency-mask width $F = 6$, and a time-mask width $T = 8$.

\section{Experimental Results}
\label{sec:experiments}

We evaluate EdgeSpot against BC-ResNet baselines under the same prototype-based FS-KWS protocol as our prior work~\cite{Our_FSKWS}. For fairness, we adapt BC-ResNet to the FS-KWS setting by replacing the final classifier to produce 64-dimensional embeddings compatible with our KD setup.

Table~\ref{tab:edge_mswc_gsc_shots} summarizes 1-shot and 10-shot results on MSWC and GSC. For GSC, we used 100 random trials with K-shot enrollments of the 11 target keywords (including silence) against the remaining 25 words. For MSWC, we utilize 3 independent runs, and each run consists of 100{,}000 positive and negative examples each.

On MSWC, we evaluate performance using DET@X, defined as the detection rate at a FAR of X. EdgeSpot consistently outperforms the re-trained BC-ResNet baselines under the same FS-KWS protocol, with the largest gains at low FARs and in the 1-shot setting. EdgeSpot-4 closely follows the Teacher, particularly on DET@5\% and AUROC, with only a small gap at DET@1\% in both 1-shot and 10-shot settings. Moreover, EdgeSpot-4 surpasses ResNet15 on nearly all metrics.

Cross-domain tests on GSC further highlight the benefits of the proposed architecture. EdgeSpot improves over BC-ResNet at all scales, most notably at FAR=1\%, and EdgeSpot-4 outperforms ResNet15 on all metrics. With 10-shot enrollment, EdgeSpot-4 attains near-Teacher performance and even exceeds it at more relaxed operating points. In the 1-shot case, the Teacher remains best, yet EdgeSpot-4 narrows the gap substantially relative to BC-ResNet, demonstrating strong transfer despite distribution shift.

\section{Discussion}
\label{sec:discussion}
We evaluated EdgeSpot under the same prototype-based FS-KWS protocol as our prior work~\cite{Our_FSKWS} and, for fairness, re-trained BC-ResNet baselines with the identical KD+SCAF loss tailored to FS-KWS. EdgeSpot delivers higher low-FAR accuracy than BC-ResNet on both MSWC and cross-domain GSC with minimal additional on-device cost. Notably, EdgeSpot-4 outperforms the ResNet15 baseline accuracy with fewer computational requirements, and nearly matches the SSL Teacher across all metrics, especially on GSC in the 10-shot condition. On MSWC, performance at relaxed operating points is effectively saturated relative to the Teacher. In the more challenging 1-shot regime, the Teacher remains strongest as expected, yet EdgeSpot-4 substantially narrows the gap while consistently outperforming BC-ResNet.

Overall, these findings demonstrate that EdgeSpot, when distilled with SCAF, recovers much of the Teacher’s discriminative power at strict FARs, while preserving an edge-friendly resource footprint.

%\vfill\pagebreak

\bibliographystyle{IEEEbib}
\bibliography{references}

\end{document}